\documentclass[submission,copyright,creativecommons]{eptcs}
\providecommand{\event}{FROM 2023} 

\usepackage{iftex}
\usepackage{amsmath}
\usepackage{amssymb}
\usepackage{amsthm}
\usepackage{xcolor}
\usepackage{MnSymbol}
\usepackage{stmaryrd}
\usepackage{xspace}
\usepackage{listings}
\usepackage[scr]{rsfso}
\usepackage{mathpartir} 
\usepackage{prftree}
\usepackage{graphics}
\usepackage[inline]{enumitem}

\usepackage{hyperref}
\usepackage{tikz-cd}

\newcommand{\limplies}{\rightarrow}
\newcommand{\lequiv}{\leftrightarrow}

\newcommand{\inh}[1]{\ensuremath{\top_{\!\!#1}}}

\newcommand{\cln}{{\kern0.2ex:\kern0.2ex}}
\newcommand{\ld}{\mathord{.\;}}

\newcommand\sbullet[1][.5]{\mathbin{\vcenter{\hbox{\scalebox{#1}{$\bullet$}}}}}

\newcommand{\EV}{\mathit{EV}}
\newcommand{\SV}{\mathit{SV}}

\newcommand{\pset}[1]{\mathbb{P}(#1)}

\newcommand{\floor}[1]{\left\lfloor#1\right\rfloor}
\newcommand{\prule}[1]{\scalebox{.95}[1]{\textnormal{{(#1)}}}}
\newcommand{\pto}{\rightharpoonup}

\newcommand{\ev}[2]{\left|#1\right|_{#2}}

\newcommand{\evMrho}[2]{\left|#1\right|_{#2,\rho}}

\newcommand{\apd}{\mathbin{\sbullet}}
\newcommand{\uapdu}{\_{\apd}\_}
\newcommand{\ceil}[1]{\lceil#1\rceil}
\newcommand{\pr}[1]{\left\langle #1 \right\rangle}
\newcommand{\smbceil}{\mathsf{def}}
\newcommand{\oto}{\mathbin{\textnormal{\textcircled{\small$\to$}}}}
\newcommand{\soto}{\mathbin{\textnormal{\textcircled{\scriptsize$\to$}}}}

\newcommand{\nil}{\mathit{nil}}

\newcommand{\cons}{\mathit{cons}}

\newcommand{\hd}{\mathit{hd}}
\newcommand{\tl}{\mathit{tl}}
\newcommand{\out}{\mathit{out}}
\newcommand{\nxt}{\mathit{nxt}}

\newcommand{\List}{\mathit{List}}

\newcommand{\Sort}{\mathit{Sort}}

\newcommand{\Sorts}{\inh{\mathit{Sort}}}

\newcommand{\Id}{\mathit{ID}}

\newcommand{\DEF}{\mathsf{EQUALITY}}
\newcommand{\SORTS}{\textnormal{\textsf{SORT}}}
\newcommand{\INH}{\mathsf{SORT}}
\newcommand{\FUN}{\mathsf{FUN}}

\newcommand{\PAIR}{\mathsf{PAIR}}

\newcommand{\Smbs}{{\color{blue}\small\textsf{Symbols}:}}
\newcommand{\Axms}{{\color{blue}\small\textsf{Axioms}:}}
\newcommand{\Sugar}{{\color{blue}\small\textsf{Notations}:}}

\newcommand{\Impts}{{\color{blue}\small\textsf{Imports}:}}
\newcommand{\axname}[1]{\textnormal{(\textsc{#1})}}
\newcommand{\sn}[1]{\mathbb{n}\;{#1}}

\lstdefinelanguage{matchinglogic}
{morekeywords={spec, endspec, theory, endtheory, model, endmodel, of},
sensitive=false,
morecomment=[l][\textit]{//},
morecomment=[s]{/*}{*/},
mathescape=true,
escapeinside={(*}{*)},
frame=none,
xleftmargin=0ex,
xrightmargin=0ex,
aboveskip=0ex,
belowskip=0ex,
basicstyle=\ttfamily\small,        
keywordstyle=\color{blue},       
}

\theoremstyle{plain}

\theoremstyle{definition}
\newtheorem{defn}{Definition}[section]
\newtheorem{exmp}{Example}[section]

\theoremstyle{remark}
\newtheorem*{rem}{Remark}

\newcommand{\catC}{\mathbb C}
\newcommand{\objZero}{\mathbf 0}
\newcommand{\objOne}{\mathbf 1}
\newcommand{\initMorph}{\raisebox{2pt}{\hbox{!`}}\xspace}
\newcommand{\initMorphIt}{\mathit{iniMor}\xspace}
\newcommand{\finMorph}{\hbox{!}\xspace}
\newcommand{\finMorphIt}{\mathit{finMor}\xspace}

\ifpdf
  \usepackage{underscore}         
  \usepackage[T1]{fontenc}        
\else
  \usepackage{breakurl}           
\fi

\title{Matching-Logic-Based Understanding of Polynomial Functors and their Initial/Final Models}
\author{Dorel Lucanu
\institute{Alexandru Ioan Cuza University of Iași, Romania}
\email{dorel.lucanu@gmail.com}
}
\def\titlerunning{Matching-Logic-Based Understanding of of Polynomial Functors}
\def\authorrunning{D. Lucanu}

\begin{document}
\maketitle

\begin{abstract}
In this paper, we investigate how the initial models and the final models for the polynomial functors can be uniformly specified in matching logic.
\end{abstract}

\section{Introduction}

It is known that many data types used in programming are defined as initial algebra or final coalgebra  for an appropriate functor $F:\catC\to \catC$, where $\catC$ is a category of data types. In this paper we assume that $\catC$ is the category of sets (see, e.g.,~\cite{Adamek03}).
If $F$ is bicontinous, i.e., it preserves the colimits of $\omega$-sequences and and the limits of $\omega^\mathit{op}$-sequences, then the initial algebra (model) is obtained via colimit of the $\omega$-sequence
\[
\objZero \xrightarrow{\initMorph} F\;\objZero \xrightarrow{F\;\initMorph} F\;F\;\objZero=F^2\;\objZero \xrightarrow{F^2\;\initMorph} F^3\;\objZero \xrightarrow{F^3\;\initMorph} \cdots \tag{\sc Ini}\label{eq:ini}
\]
where $\objZero$ is the initial object in $\catC$, and $\objZero \xrightarrow{\initMorph} X$ is the unique arrow from the initial object, and the final coalgebra (model) is the limit of the $\omega^\mathit{op}$-sequence
\[
\objOne \xleftarrow{\finMorph} F\;\objOne \xleftarrow{F\;\finMorph} F\;F\;\objOne=F^2\;\objOne \xleftarrow{F^2\;\finMorph} F^3\;\objOne \xleftarrow{F^3\;\finMorph} \cdots \tag{\sc Fin}\label{eq:fin}
\]
where $\objOne$ is the final object in $\catC$, and $\objOne \xleftarrow{\finMorph}  X$ is the unique arrow to the final object~\cite{Adamek03}.

This is a nice abstract framework, but, as we know, the evil is hidden in details.  How the elements of the initial and final models look like for various concrete functors? How could they be handled in practice?

A possible answer can be obtained by capturing these objects in Matching Logic (ML), the logical foundation of the K Framework, where the program languages and the properties of their programs can be specified in a uniform way (see, e.g., ~\cite{Ros17,CR19,Rosu020,CLR21a,LinCTWR23}). First steps are done in~\cite{CLR20a}, where the initial algebra semantics is captured in ML, and in~\cite{CLR21a}, where it is shown how examples of inductive/coinductive data types are fully specified in ML.
We say that ML captures a (inductive/coinductive) data type $\mathit{DT}$ if there is an ML theory $\mathsf{Th}^{\sf ML}(\mathit{DT})$ such that:
\begin{itemize}
\item from each $\mathsf{Th}^{\sf ML}(\mathit{DT})$-model $M$ we may extract a structure $\alpha(M)$ that is isomorphic to $\mathit{DT}$, and
\item each deduction principle for $\mathit{DT}$ (e.g., induction or coinduction) can be expressed as a theorem within ML using its proof system. 
\end{itemize}

In this paper, we investigate how data types specified as initial $F$-algebras or as a final $F$-coalgebras, where $F$ is a polynomial functor, can be captured in ML. 
The polynomial functors can be defined in two ways:
\begin{enumerate}
\item Using "classical" inductive definition of polynomials (see, e.g., \cite{rutten00}): the polynomial functors is the smallest class including the constant and the identity functors, and it is closed to sum, product, and constant-exponent functors.
\item Using unary container functors (see, e.g.,\cite{altenkirchGHMM15}): a polynomial functor is of the form $X\mapsto \sum_{a:A}X^{B[a]}$, where $a:A\vdash B[a]$ is an $A$-indexed family.
\end{enumerate}

The constant and the identity functors, together with their initial and final models, can be easily captured in ML. Moreover, if we exclude the exponent functor, then the initial algebra can be captured using the approach similar to that from~\cite{CLR20a}. The exponent functor complicates the things.
A possible approach for the classical definition is as follows: supposing that we have captured $F_1\;X$ and $F_2\;X$ by the ML theories (specifications) $\mathrm{SPEC}(F_1\;X)$ and resp. $\mathrm{SPEC}(F_2\;X)$, then use these specifications to build $\mathrm{SPEC}(F_1\;X~\mathit{op}~F_2\;X)$, i,.e., to obtain something like 
\[
\mathrm{SPEC}(F_1\;X~\mathit{op}~F_2\;X) = \mathrm{SPEC}(F_1\;X)~\overline{\mathit{op}}~\mathrm{SPEC}(F_2\;X)
\]
where $\overline{\mathit{op}}$ reflects $\mathit{op}$ to the level of ML specifications, and it follows to be defined.
In order to accomplish that, we need a "uniform standard" definition for the specifications $\mathrm{SPEC}(F\;X)$.

The container functors already have a uniform standard definition, and therefore they are more tempting for our investigation.
This approach is a work-in-progress, and the results obtained up now show that:
\begin{itemize}
\item it is possible to specify unary container functors, together with their initial algebras and final coalgebras, as ML theories;
\item it is possible to derive the induction principle and the coinduction principle as theorems of the corresponding theories;
\item the ML reasoning can be used to understand better the intimate structure of the initial algebra and the final coalgebra.
\end{itemize}
The approach is instantiated on the lists example, in order to see the relationship with the classical approach of these data types, and on that of Moore machines, in order to see the (constant) exponential functor at work.

The paper is structured as follows. Section~\ref{sec:pf} recalls the definitions for polynomial functors and for unary container functors, and the the relationship between them.
Section~\ref{sec:ml} briefly recalls the main elements of matching logic, together with the theories of the equality and of the sorts.
Section~\ref{sec:fc-ml} includes the main contribution, showing that how the unary container functors and their related concepts can be specified in matching logic. The instantiation of the general approach on the examples of lists and Moore machines are included in Section~\ref{sec:lists} and Section~\ref{sec:Moore}, respectively. The paper ends with some concluding remarks.

\section{Polynomials Functors}
\label{sec:pf}

This section briefly recalls the definition of the polynomial functors and their (co)algebras.
We consider only the particular case of the functors defined over the category of sets $\catC$.

\begin{defn}\cite{jacobsRutten2012Introd}
Given a functor $F: \catC\to \catC$, an \emph{$F$-algebra} consists of an object $X$ in $\catC$ and an arrow $\alpha: F\;X\to X$.
An algebra morphism $(X,\alpha)\to (X',\alpha')$ is an arrow $h:X\to X'$ in $\catC$ such that $h\circ \alpha=\alpha'\circ F\;h$.
An \emph{initial algebra} for the functor $F$ is an initial object in the category of $F$-algebras and $F$-algebra
morphisms.
\end{defn}

\begin{exmp}[Lists]\label{ex:lists}
The lists over a set of elements $E$ can be defined as an $L$-algebra $\alpha: L\; X\to X$ for the functor $L:\catC\to \catC$ given by $L\; X = \objOne + E \times X$.
Usually, such an algebra is defined by a constant $\nil:\objOne\to X$ and a binary operation $\cons: E\times X\to X$. The initial $L$-algebra is isomorphic to the finite lists inductively defined~\cite{sangiorgi-bisim-coind} by the grammar
\begin{align}
\List ::= nil \mid \cons(E, \List) \label{bnf:lists}\tag{\textsc{Lst}}
\end{align}
We use $\mu X\ld L\;X$ or $\mu L$ to denote the initial model of the functor $L$. 
\end{exmp}

\begin{defn}\cite{jacobsRutten2012Introd}
Given a functor $F: \catC\to \catC$, an \emph{$F$-coalgebra} consists of an object $X$ in $\catC$ and an arrow $\gamma: X\to F\;X$.
A coalgebra morphism $(X,\gamma)\to (X',\gamma')$ is an arrow $h:X\to X'$ in $\catC$ such that $h\circ \gamma=\gamma'\circ F\;h$.
An \emph{final coalgebra} for the functor $F$ is a final object in the category of $F$-coalgebras and $F$-coalgebra
morphisms.
\end{defn}

\begin{exmp}[Colists]
The colists over a set of elements $E$ can be defined as an $L$-coalgebra $\gamma: X\to L\; X$, where $L$ is the functor used for lists (see Example~\ref{ex:lists}).
Usually, such a coalgebra is defined by
 \begin{itemize}
\item  a total operation ${\_}\ld\!\! ?:X\to \{1,2\}$ such that $x\ld\!\! ? = \mathsf{if}~g(x)=\iota_1(\star)~\mathsf{then}~1~\mathsf{else}~2~\mathsf{fi}$, where $\star$ is the unique element of $\objOne$, and 
\item two partial operations:  ${\_}\ld\hd: X\pto E$, and ${\_}\ld\tl: X\pto X$ such that $x\ld\hd=e$ and $x\ld\tl=x'$  iff $x\ld\!\! ?=2$ and $g(x)=\iota_2(\pr{e,x'})$. 
\end{itemize}

The final coalgebra is isomorphic to the possible infinite lists coinductively defined by the grammar~\eqref{bnf:lists}~\cite{sangiorgi-bisim-coind}.
We use $\nu X\ld L\;X$ or $\nu L$ to denote the final model of the functor $L$.
\end{exmp}

\begin{exmp}[Moore machines]
A Moore machine is an $M$-coalgebra $\alpha: X\to M\;X$, where $M$ is the functor $M: X \to O \times X ^I$. Usually, a Moore machine is defined by an \emph{output} function $\mathit{out}: X\to O$ and a \emph{transition}  function $\mathit{tr}:X\to X^I$~\cite{rutten00}.
The final coalgebra $\nu X\ld M\;X$ is isomorphic to $(\overline{\mathit{out}},\overline{\mathit{tr}}): O^{I^*}\to O\times (O^{I^*})^I$, $\overline{\mathit{out}}(f)=f(\varepsilon)$, $\overline{\mathit{tr}}(f)=g$ with $g(i)(w)=f(\langle i\rangle\cdot w)$ for $i\in I$ and $w\in I^*$, where $f$ is a function $F: I^*\to O$, $g$ a function $g:I \to O^{I^*}$, $\varepsilon$ the empty sequence, and $\cdot$ the concatenation of sequences.
\end{exmp}

\begin{defn}\cite{rutten00}
The class of \emph{polynomial functors} is inductively defined as follows: 
\begin{itemize}
\item the constant functor $A$ (where $A$ is an object in $\catC$) is a polynomial functor; 
\item the identity functor $\Id$  is a polynomial functor; 
\item the sum $F_1+F_2$ of  two polynomial functors $F_1$ and $F_2$  is a polynomial functor; 
\item the product $F_1 \times F_2$ of  two polynomial functors $F_1$ and $F_2$  is a polynomial functor;  and 
\item the function space functor $F(X)=X^A$, where A is an arbitrary object.
\end{itemize}
\end{defn}

An alternative to define polynomial functors is given by the (unary) container functors \cite{abbottAG05,altenkirchM09,altenkirchGHMM15}.

\begin{defn}
An \emph{$A$-indexed family $a:A \vdash B[a]$} is a family of objects of $\catC$ indexed by elements of $A$. Categorically, it is an object $B$ of $\catC/A$ and $B[a]$ denotes the elements of $B$ mapped to $a$. 
\end{defn}

\begin{defn}
Given an $A$-indexed family $a:A \vdash B[a]$, the \emph{dependent product} $\prod_{a\cln A}B[a]$ is the object of the \emph{dependent functions}, which maps an $a\cln A$ into a $b\cln B[a]$. Set theoretically, we have
\[
\prod_{a\cln A}B[a]=\left\{f\in (\bigcup_{a\cln A}B[a])^A \middle\mid \forall a\cln A. f(a)\in B[a]\right\}  
\]
The \emph{dependent sum} $\sum_{a\cln A}B[a]$ is the dual of the dependent product and it consists of the pairs $\pr{a,b}$ with $a\cln A$ and $b\cln B[a]$.
\end{defn}

\begin{defn}
A functor $F : \catC \to \catC$ is a \emph{ (unary) container functor} iff it is naturally isomorphic
to a functor of the form $F\; X =  \sum_{a\cln A} X^{B[a]}$, for some objects $A$ in $\catC$ and an $A$-indexed family $a\cln A \vdash B[a]$.
\end{defn}

\begin{rem}
An element of $\sum_{a\cln A} X^{B[a]}$ is a pair $\pr{a, f}$, where $a \cln A$ is the \emph{shape} and $f: B[a] \to X$ is the function that labels the \emph{positions} $B[a]$ with elements from $X$. 
Another way to define $B$ is as an object in $\catC/A$.
\end{rem}

\subsection{Polynomial Functors as Container Functors}

Here we recall the relationship between polynomial functors and container functors (see, e.g., ~\cite{abbottAG05}).

\subsubsection*{Constant Functor} 
The main idea is to identify the constant value with the shapes $A$.
Consider $B$ as being $a\cln A\vdash \objZero$ (no positions of shape $a$),  where $\objZero$ denotes the initial object of $\catC$. 
We get  $\sum_{a\cln A} X^{B[a]} \approxeq \sum_{a\cln A} X^{\objZero} \approxeq A$.
\begin{rem}
The elements of $ \sum_{a\cln A} X^{\objZero}$ are pairs $\pr{a, f: \objZero\to X}$, where $a\in A$. Since $f: \objZero\to X$ is unique, we obtain $\pr{a, f: \objZero\to X}\approxeq a$.
\end{rem}

\subsubsection*{Identity Functor}
Consider $A$ as being $\objOne$ (just one shape) and $B$ as being $\star\cln \objOne\vdash\objOne$ (just one position), where $\star$ is the unique element in $\objOne$. It follows that
$\sum_{a\cln A} X^{B[a]} \approxeq \sum_\star X^\objOne\approxeq X$.
\begin{rem}
The elements of $\sum_\star X^\objOne$ are pairs $\pr{\star, f: \objOne\to X}$. Since $f$ selects just one element $x$ in $X$, it follows that $\pr{\star, f: \objOne\to X}\approxeq x$.
\end{rem}

\subsubsection*{Sum}
Assume that $F\;X = \sum_{a\cln A} X^{B[a]}$ and $F'\;X = \sum_{a'\cln A'} X^{B'[a]'}$. Then $(F+F')\;X$ is  $\sum_{a''\cln A+A'} X^{[B,B'][a'']}$, where $[B,B'][a'']=B[a]$ if $a''=\iota_1\;a$, and $[B,B'][a'']=B'[a']$ if $a''=\iota_2\;a'$.
\begin{rem}
 The following commutative diagram may help to understand the definition of $a''\cln A+A'\vdash [B,B'][a'']$:
\begin{center}
\begin{tikzcd}
 & A+A'  & \\
A \arrow[ur,"\iota_1"] && A'  \arrow[ul, "\iota_2"']\\
B \arrow[u]  \arrow[r] & \hbox{$B + B'$} \arrow[uu]  & B' \arrow[u] \arrow[l]
 \end{tikzcd}
\end{center}
The arrow $B+B'\to A+A'$ is equivalently written as the $A+A'$-indexed set $a''\cln A+A'\vdash [B,B'][a'']$.
Set theoretically,  $\sum_{a''\cln A+A'} X^{[B,B'][a'']}$ is the set of pairs $\pr{a'', [f, f']: [B,B'']\; a'' \to X}$, where 
\begin{itemize}
\item either $a''=\iota_1[a]$, $\pr{a, f:B[a]\to X}$ in $F\;X$, and $\forall x\cln B[a]\ld [f, f']\; x= f\;x$, or
\item   $a''=\iota_2[a]'$, $\pr{a',f':B'[a]'\to X}$ in $F'\;X$, and $\forall x'\cln B[a']\ld [f, f']\; x'= f'\;x'$.
\end{itemize}
 In other words, the shape of the sum is the sum of the component shapes, and a labelling function for the sum is the sum of two corresponding component labelling functions. 
\end{rem}

\subsubsection*{Product}
We have $(F+F')\;X =  \sum_{\pr{a,a'}\cln A\times A'} X^{\langlebar B,B'\ranglebar[\pr{a,a'}]}$, where $F\;X$ and $F'\;X$ are similar to those from the sum, and $\langlebar B, B'\ranglebar[\pr{a,a'}]=B[a] + B'[a']$ (the positions of the product is the disjoint union of the component positions). 
\begin{rem}
We prefer to write $\langlebar B, B'\ranglebar[a,a']$ for  $\langlebar B, B'\ranglebar[\pr{a,a'}]$.
Set theoretically,  $\sum_{\pr{a.a'}\cln A\times A'} X^{\langlebar B, B'\ranglebar[a,a']}$ is the set of pairs $\pr{\pr{a,a'}, [f, f']: B[a]+B'[a'] \to X}$, where $f:B[a]\to A$ and $f':B'[a']\to A'$. In other words, the shape of the product is the product of the component shapes and a labelling function for the product is the sum of two corresponding component labelling functions. 
\end{rem}

\subsubsection*{Exponentiation}
Assuming $F\;X = \sum_{a\cln A} X^{B[a]}$, the (constant) exponent functor $(F\;X)^C$ is 
$\sum_{g:C\to A} X^{\sum_{c\cln C}B[g\;c]}$.
\begin{rem}
An element of $\sum_{g:C\to A} X^{\sum_{c\cln C}B[g\;c]}$ is a pair $\pr{g, f}$ consisting of a function $g:C\to A$ assigning  shapes to $C$ and a $C$-indexed function $c\cln C\vdash f_c: B[g\;c]\to X$ labelling the positions $B[g\;c]$ for each $c$ in $C$.
\end{rem}

\section{Matching Logic (ML)}
\label{sec:ml}

 Matching logic \cite{Ros17,CR19,CLR21a} provides a unifying framework for defining semantics of programming languages. 
A programming language is defined in 
matching logic as a \emph{logical theory}, i.e., a set of axioms.
The key concept in matching logic is that of \emph{patterns},
which can be \emph{matched} by certain elements. 
By building complex patterns, we can match elements that have
complex structures or certain properties, or both. 
The presentation of matching logic in this review section 
follows \cite{CLR21a}. 

\begin{defn}
\label{def:AML-syntax}
Let us fix two sets $\EV$ and $\SV$.
The set $\EV$ includes \emph{element variables} $x,y,\dots$.
The set $\SV$ includes \emph{set variables} $X,Y,\dots$.
A \emph{matching logic signature} $\Sigma$ is a set of \emph{(constant) symbols}, denoted $\sigma,\sigma_1,\sigma_2,\dots$.
Let us fix a signature $\Sigma$.
The set of \emph{($\Sigma$-)patterns} is inductively defined as follows:
\begin{align*}\label{eq:MLsyntax}
\varphi ::= 
& \ x 
\mid X 
\mid \sigma
\mid \varphi_1 \  \varphi_2
\mid \bot
\mid \varphi_1 \limplies \varphi_2
\mid \exists x \ld \varphi
\mid \mu X \ld \varphi 
\end{align*}
where in $\mu X \ld \varphi$, called a least-fixpoint pattern,
we require that $\varphi$ is positive in $X$,
i.e., $X$ does not occur in an odd number 
of times of the left-hand sides of implications $\varphi_1 \limplies \varphi_2$. 
\end{defn}

\begin{defn}
\label{def:AML-models}
A \emph{(matching logic) $\Sigma$-model} $M$ consists of
\begin{enumerate}
\item a nonempty \emph{carrier set}, which we also denote $M$; 
\item an \emph{application function} $\uapdu \colon M \times M \to 
\pset{M}$,   where $\pset{M}$ is the powerset of $M$; and
\item a \emph{symbol interpretation} $\sigma_M \subseteq M$ as a subset
     for $\sigma \in \Sigma$.
\end{enumerate}
\end{defn}

\begin{defn}
\label{def:AML-semantics}
Given $M$ and 
a \emph{variable valuation} $\rho \colon (\EV \cup \SV) \to  M 
\cup \pset{M}$ such that
$\rho(x) \in M$ for $x\in \EV$ 
and $\rho(X) \subseteq M$ for $X \in \SV$,
we inductively define \emph{pattern valuation} $\evMrho{\varphi}{M}$
as follows:
\begin{enumerate}
	\item $\evMrho{x}{M} = \{ \rho(x) \}$ for  $x \in \EV$ 
	\item $\evMrho{X}{M} = \rho(X)$ for  $X\in \SV$
	\item $\evMrho{\sigma}{M} = \sigma_M$ for $\sigma \in \Sigma$
	\item $\evMrho{\varphi_1 \, \varphi_2}{M} = 
	\bigcup_{a_i \in \evMrho{\varphi_i}{M},i \in \{1,2\}} a_1 \apd a_2$;
	note that $a_1 \apd a_2$ is a subset of $M$.
	\item $\evMrho{\bot}{M} = \emptyset$
	\item $\evMrho{\varphi_1 \limplies \varphi_2}{M} = M \setminus 
	(\evMrho{\varphi_1}{M} \setminus \evMrho{\varphi_2}{M})$
	\item $\evMrho{\exists x \ld \varphi}{M} = \bigcup_{a \in M} 
  \ev{\varphi}{M,\rho[a/x]}$
  \item $\evMrho{\mu X \ld \varphi}{M} = \mathbf{lfp} (A \mapsto \ev{\varphi}{M, \rho[A/X]})$
\end{enumerate}
where
$\rho[a/x]$ (resp. $\rho[A/X]$) is the valuation $\rho'$
with $\rho'(x) = a$ (resp. $\rho'(X) = A$) and
agreeing with $\rho$ on all the other variables.
 in $\EV \cup 
\SV \setminus \{x\}$ (resp. $\EV \cup \SV \setminus \{X\}$).
We use $\mathbf{lfp} (A \mapsto \ev{\varphi}{M, \rho[A/X]})$ to denote the 
smallest set $A$ such that $A = \ev{\varphi}{M, \rho[A/X]}$.
The existence of such an $A$ is guaranteed by the requirement
that $\varphi$ is positive in $X$.
We abbreviate $\ev{\varphi}{M,\rho}$ as $\ev{\varphi}{\rho}$
and further as $\ev{\varphi}{}$ if $\varphi$ is closed.
\end{defn}

\begin{defn}
\label{def:valid}
We say that $\varphi$ \emph{holds} in $M$, written
$M \vDash \varphi$,
if $\ev{\varphi}{M,\rho} = M$ for all $\rho$.
For a pattern set $\Gamma$,  
we write $M \vDash \Gamma$, if $M \vDash \psi$ for all $\psi \in \Gamma$.
We write $\Gamma \vDash \varphi$, if $M \vDash \Gamma$ implies
$M \vDash \varphi$  for all $M$. 
\end{defn}

The following common constructs can be defined from the basic pattern syntax
as syntactic sugar in the usual way:
$$
\begin{aligned}
 \neg \varphi &\equiv \varphi \limplies \bot \qquad
 &\varphi_1 \vee \varphi_2 &\equiv \neg \varphi_1 \limplies \varphi_2 \qquad
 &\varphi_1 \wedge \varphi_2 &\equiv \neg(\neg \varphi_1 \vee \neg \varphi_2) \\
 \top &\equiv \neg \bot 
 &\forall x \ld \varphi &\equiv \neg \exists x \ld \neg \varphi
 &\nu X \ld \varphi &\equiv \neg \mu X \ld \neg \varphi[\neg X / X]
\end{aligned}
$$
We assume the standard precedence between the above constructs.

\subsubsection*{Equality}
The equality can be defined as a derived construct (see~\cite{Ros17,CLR21a}).
Two patterns $\varphi_1$ and $\varphi_2$ are equal, 
written $\varphi_1 = \varphi_2$, iff
it is equivalent to $\top$ if the
two patterns are matched by the same elements.
Otherwise, it is equivalent to $\bot$. To express that in ML,
a new symbol $\mathit{def} \in \Sigma$ is introduced, called the \emph{definedness} symbol,
and specify it with the axiom $\prule{Definedness}$. The resulted theory can be described as follows:
\begin{lstlisting}[language=matchinglogic, basicstyle=\linespread{1.2}\ttfamily\small]
spec $\DEF$
  (*\Smbs*)  $\smbceil$
  (*\Sugar*) $\ceil{\varphi} \equiv \smbceil \  \varphi$
  (*\Axms*)   $\prule{Definedness} \quad \forall x \ld \ceil{x} $
  (*\Sugar*) 
    $\floor{\varphi} \equiv \neg \ceil{\neg \varphi}
    $                 (*// totality*)
    $\varphi_1 = \varphi_2 \equiv \floor{\varphi_1 \lequiv \varphi_2}
    $                 (*// equality*)
    $\varphi_1 \subseteq \varphi_2 \equiv \floor{\varphi_1 \limplies \varphi_2}
    $                 (*// set inclusion*)
    $x \in \varphi \equiv x \subseteq \varphi
    $                 (*// membership*)
endspec
\end{lstlisting}

\subsubsection*{Sorts}
Matching logic has no builtin support for sorts.
Instead, we define a \emph{theory of sorts}
to support arbitrary sort structures
following the ``sort-as-predicate'' paradigm.
A \emph{sort} has a name and 
is associated with a set of its \emph{inhabitants}. 
In matching logic, we use a symbol $s \in \Sigma$ to represent the sort name
and use $(\mathsf{inh} \ s)$ to represent all its inhabitants,
where $\mathsf{inh} \in \Sigma$ is an ordinary symbol.
For better readability, we define
the notation $\inh{s} \equiv \mathsf{inh} \ s$. 
\begin{lstlisting}[language=matchinglogic, basicstyle=\linespread{1.1}\ttfamily\small]
spec $\INH$  (*\Impts*) $\DEF$
  (*\Smbs*) $\mathit{inh},\Sort$
  (*\Sugar*)
   $\inh{s} \equiv \mathit{inh} \  s
   $                              (*// inhabitants of sort $s$*) 
   $s_1\le s_2\equiv \inh{s_1} \subseteq \inh{s_2}
   $                              (*// subsort relation*)
   $\neg_s \varphi \equiv (\neg \varphi) \wedge \inh{s}
   $                              (*// negation within sort $s$*) 
   $\forall x \cln s \ld \varphi \equiv \forall x \ld x \in \inh{s} \limplies 
 \varphi
   $                              (*// $\forall$ within sort $s$*) 
   $\exists x \cln s \ld \varphi \equiv \exists x \ld x \in \inh{s} 
 \land 
 \varphi
   $                              (*// $\exists$ within sort $s$*) 
   $\mu X \cln s \ld \varphi \equiv \mu X \ld X \subseteq \inh{s} 
 \land 
 \varphi
   $                              (*// $\mu$ within sort $s$*)
   $\nu X \cln s \ld \varphi \equiv \nu X \ld X \subseteq \inh{s} 
 \land 
 \varphi
   $                              (*// $\nu$ within sort $s$*)
   $\varphi \cln s \equiv \exists z \cln s \ld \varphi = z
   $                              (*// \textrm{``}typing\textrm{''}*) 
   $f\cln s_1\otimes\cdots\otimes s_n\to s\equiv
 \forall x_1\cln s_1\ldots\forall x_n\cln s_n\ld \exists y\cln s\ld f\, x_1\ldots x_n = y$
                               (*// functional*) 
   $f\cln s_1\otimes\cdots\otimes s_n\pto s\equiv
 \forall x_1\cln s_1\ldots\forall x_n\cln s_n\ld \exists y\cln s\ld f\, x_1\ldots x_n \limplies y$
                               (*// partially functional*)
  (*\Axms*) 
   $\exists x\ld \Sort = x$
   $\Sort\in\Sorts$
endspec
\end{lstlisting}

The ML specifications for the sum, product, and function sorts are given in \cite{CLR21a}. For reader convenience, we recall them in Appendix~\ref{ap:sorts}. 

\section{Specifying Initial Algebra and Final Coalgebra for Container Functors in ML}
\label{sec:fc-ml}

In this section we show how to specify a unary container functor in ML and how to extract the structures for their initial algebra and the final coalgebra.

\subsection{Capturing Elements from the Category $\catC$}

\begin{itemize}
\item $\objZero$ is specified by $\bot$;
\item $\objOne=\{\star\}$ is specified by
\begin{itemize}
\item  a symbol $\mathit{star}\in \Sigma$; 
\item a notation: $\star \equiv \mathit{star}$; and
\item an axiom\\
$~~~~\begin{aligned}
&\exists y\ld \star = y  && \axname{singleton}\\
\end{aligned}$
\end{itemize}
\item $\objZero\xrightarrow{\initMorph} X$ is specified by
\begin{itemize}
\item  a symbol $\initMorphIt\in \Sigma$; 
\item a notation: $\initMorph \equiv \initMorphIt$; and
\item an axiom\\
$~~~~\begin{aligned}
&\forall x \ld \initMorph\; x = \bot  && \axname{\textrm{captures~}$\objZero\xrightarrow{\initMorph} X$}\\
\end{aligned}$
\end{itemize}
\item $\objOne\xleftarrow{\finMorph} X$ is specified by
\begin{itemize}
\item  a symbol $\finMorphIt\in \Sigma$; 
\item a notation: $\finMorph \equiv \finMorphIt$; and
\item an axiom\\
$~~~~\begin{aligned}
&\forall x \ld \finMorph\; x = \star  && \axname{\textrm{captures~}$\objOne\xleftarrow{\finMorph} X$}\\
\end{aligned}$
\end{itemize}
\end{itemize}
\begin{rem}
An alternative way to specify $\objOne$ is by $\top$, in which case $\objOne\xleftarrow{\finMorph} X$ is the inclusion: $\forall x\ld x\in X \limplies \finMorph\;x=x$. This version is used when we compute the greatest fixpoint.
\end{rem}

\subsection{Expressing Indexed Families in ML}

An $A$-indexed family $a\cln A \vdash B[a]$ is specified by a constant symbol $\mathit{depOf}\in\Sigma$, a notation $B[a]\equiv \mathit{depOf}\;B\;a$, and two axioms:
\begin{itemize}
\item $A$ is a sort: $A\cln \Sort$ (equivalent to $A\in\Sorts$), and 
\item for each $a\cln A$, $B[a]$ is a sort: $\forall a \cln A\ld B[a]\cln\Sort$
\end{itemize}
where we assumed that $A$ and $B$ are specified as functional patterns.

\subsection{Expressing Dependent Products/Sums in ML}

A dependent product $\prod_{a\cln A}B[a]$ is specified by a constant symbol $\mathit{DepProd}\in \Sigma$, a notation $\prod_{a\cln A}B[a] \equiv \mathit{DepProd}\;A\;B$ ($a$ plays a local role and its name can be changed\footnote{Actually, $\Pi$ should be captured as a binder~\cite{ChenR20}, but this is not needed for the purpose of this paper.}), and by adding the axioms:
\begin{itemize}
\item $\prod_{a\cln A}B[a]$ is a sort: $\prod_{a\cln A}B[a]\cln\Sort$ 
\item $\inh{\prod_{a\cln A}B[a]}$ is the set of dependent functions:
 \[
\inh{\prod_{a\cln A}B[a]}=\exists f\ld f \land ((\ceil{\inh{A}}\land \forall a\cln A\ld \exists b\cln B[a]\ld f\;a=b) \lor (\neg\ceil{\inh{A}}\land f=\initMorph))
 \]
 The above axiom distinguishes between two cases: the sort $A$ is non-empty, in which case $ \inh{\prod_{a\cln A}B[a]}$ includes the dependent functions that maps an $a\cln A$ into a $b\cln B[a]$, and  when the sort $A$ is empty, in which case $\inh{\prod_{a\cln A}B[a]}$ consists of the function given by the initial morphism. 
\end{itemize}
\noindent
Similarly, a dependent sum $\sum_{a\cln A}B[a]$ is specified by a constant $\mathit{DepSum}\in \Sigma$, a notation $\sum_{a\cln A}B[a] \equiv \mathit{DepSum}\;A\;B$ (again, $a$ plays a local role and its name can be changed), and by adding the axioms:
\begin{itemize}
\item $\sum_{a\cln A}B[a]$ is a sort: $\sum_{a\cln A}B[a]\cln\Sort$ 
\item $\inh{\sum_{a\cln A}B[a]}$ is the set fo dependent pairs: \[\inh{\sum_{a\cln A}B[a]}=\exists a\cln A\ld \exists b\cln B[a]\ld \pr{a,b}\]
\end{itemize}

\subsection{Expressing Unary Container Functors in ML}

Let $X\mapsto F\;X= \sum_{a\cln A}X^{B[a]}$ be a container functor.
Recall that $F\;X$ is the set of pairs $\pr{a,f}$ with $a\cln A$ and $f\cln X^{B[a]}$ (or, equivalently, $f\cln B[a]\to X$).
If $X$ is specified as the set of inhabitants $\inh{s}$ of a sort $s\cln \Sort$ and $\inh{B[a]}\not=\bot$ (that is equivalent to $B[a]\not\approxeq\objZero$ in $\catC$), then $X^{B[a]}$ is specified by the sort $B[a]\oto s$~\cite{CLR21a} (see also Appendix~\ref{ap:func-sort}) and the specification of $\sum_{a\cln A}X^{B[a]}$ is a particular case of dependent sum specification.
Otherwise, we have to explicitly specify $X^{B[a]}$ by the notation
\[
X^{B[a]} \equiv \exists f\ld f \land (((\inh{B[a]}\not=\bot)\land\forall b\cln B[a]\ld \exists x\ld f\;b = x \land x\in X)  \lor ((\inh{B[a]}=\bot)\land f=\initMorph))
\] 
and use it directly in the specification of $\sum_{a\cln A}X^{B[a]}$:
\[
\exists a\cln A\ld \exists f\ld \pr{a, f} \land f\in X^{B[a]}
\]
which is equivalent to
\[
\exists a\cln A\ld \exists f\ld \pr{a, f} \land (((\inh{B[a]}\not=\bot)\land \forall b\cln B[a]\ld \exists x\ld f\;b = x \land x\in X) \lor ((\inh{B[a]}=\bot)\land f=\initMorph))
\]
Recall that if $\inh{B[a]}=\bot$ ($B[a]\approxeq\objZero$) then there is just one function $\objZero\xrightarrow{\initMorph} X$ specified by $\initMorph \equiv \initMorphIt$.

\subsection{Specifying Initial Algebra }

Let $X\mapsto F\;X= \sum_{a\cln A}X^{B[a]}$ be a container functor. The initial $F$-algebra is specified by using the characterization given by the "no junk and no confusion" properties for the constructors ~\cite{CLR21a}:
\begin{itemize}
\item a constructor $\cons\in\Sigma$ specified by the following axioms:\\
$\begin{aligned}
&\forall a\cln A\ld \forall f\ld f\in X^{B[a]}\limplies \exists x\ld x\in X\land\cons\;\pr{a, f} = x  && \axname{Functional}\\
&\forall a,a'\cln A\ld \forall f, f'\ld \cons\;\pr{a, f} = \cons\;\pr{a',f'} \limplies a = a' \land f = f'&& \axname{No Confusion}
\end{aligned}$
\item a sort $\mu F$ with initial semantics:\\
$\begin{aligned}
\inh{\mu F} = \mu X\ld \exists a\cln A\ld  \cons\;\pr{a,X^{B[a]}}&& \axname{No Junk}
\end{aligned}$
\end{itemize}

\noindent
\paragraph{Computing the least fixpoint} 

Let $\varphi_F(X)$ denote the pattern $\exists a\cln A\ld  \cons\;\pr{a, X^{B[a]}}$.
Given a model $M$ and a valuation $\rho$, we have  $\evMrho{\mu X \ld \varphi_F(X)}{M} = \mathbf{lfp} (A \mapsto \ev{\varphi_F(X)}{M, \rho[A/X]})$.
We also denote by $A\mapsto \phi_F(A)$ the function A $\mapsto \ev{\varphi_F(X)}{M, \rho[A/X]}$. If $\phi_F$ is continuous, then $\mathbf{lfp}(\phi_F)=\bigcup_{n\ge 0}\phi^n_F(\emptyset)=\emptyset\cup \phi_F(\emptyset)\cup \phi^2_F(\emptyset)\cup\cdots$. Since $\phi_F(\emptyset)=\ev{\varphi_F(\bot)}{M, \rho})$ and writing $\varphi(\psi)$ for $\varpi[\psi/X]$, we (informally) obtain that $\mathbf{lfp}(\phi_F)$ is the interpretation of the 
infinite disjunction
\[
\bot \lor \varphi_F(\bot) \lor \varphi^2_F(\bot) \lor \varphi^3_F(\bot) \lor \cdots \tag{\textsc{Lfp}}\label{eq:ini-ml}
\]
according to $M$ and $\rho$. We have $\varphi^n_F(\bot)\limplies \varphi^{n+1}_F(\bot)$ and each $\varphi^n_F(\bot)$ gives an approximation of $\inh{\mu F}\approxeq\mathbf{lfp}(\phi_F)$. So,  in order to understand how the elements of the initial algebra look like, we have to investigate these ML patterns.
\begin{description}
\item[ $ \varphi_F(\bot)$.] We have $X^{B[a]} =\bot ^{B[a]} \not= \bot$ iff $\inh{B[a]}=\bot$, because otherwise we have
$(\forall b\cln B[a]\ld \exists y\ld f\;b = y \land y\in \bot) =\bot$. It follows that $\bot ^{B[a]}$ consists of the unique function $\initMorph$.
We obtain 
\[ 
  \varphi_F(\bot) = \exists a_1\cln A\ld \cons\pr{a_1,\initMorph}\land (\inh{{B[a_1]}}=\bot)
\]
i.e., each $a_1\cln A$, with its corresponding dependent sort $B[a_1]$ empty, defines a constant constructor.
\\
If $\forall a\cln A\ld \inh{B[a]}\not =\bot$ then $(\exists a\cln A\ld  \cons\;\pr{a,\inh{B[a]\oto \sn{X}}})=\bot$ and hence $\inh{\mu L}=\bot$.
\item[$ \varphi^2_F(\bot)=\exists a_2\cln A\ld  \cons\;\pr{a_2, \varphi_F(\bot)^{B[a_2]}}$.] 
If $\inh{B[a_2]}=\bot$ then $\varphi_F(\bot)^{B[a_2]}$ consists of the constant constructor $\cons\;\pr{a_2,\initMorph}$, i.e., $\varphi_F(\bot)^{B[a_2]}=\bot ^{B[a_2]}$.
If $\inh{B[a_2]}\not=\bot$ then we have 
\[
\varphi_F(\bot)^{B[a_2]}=\exists f\ld f\land \forall b\cln B[a_2]\ld \exists a_1\cln A\ld f\;b=\cons\pr{a_1,\initMorph}\land (\inh{{B[a_1]}}=\bot)
\] 
We obtain
\[
\varphi^2_F(\bot) = \varphi_F(\bot) \lor \exists a_2\cln A\ld  \cons\;\pr{a_2, \varphi_F(\bot)^{B[a_2]}} \land (\inh{B[a_2]}\not=\bot)
\]
\item[\ldots]
\end{description}

\begin{rem}
The ML pattern~\eqref{eq:ini-ml} can be seen as an informal translation in ML of the colimit~\eqref{eq:ini}.
\end{rem}

\paragraph{Deriving Induction Principle}
Once we have seen how the least fixpoint is computed, we may derive the following \emph{Induction Principle}:
\[
\dfrac{\forall a\cln A\ld \cons\;\pr{a,\psi} \limplies \psi}{\inh{\mu F} \limplies \psi}
\]
The justification for this principle is similar to that for lists given in~\cite{CLR21a}.

\subsection{Specifying Final Coalgebra}

Let $X\mapsto F\;X= \sum_{a\cln A}X^{B[a]}$ be a container functor. The final $F$-coalgebra is specified by:
\begin{itemize}
\item the constructor $\cons$ together with its axioms;
\item a sort $\nu F$ with final semantics:\\
$~~~~\begin{aligned}
\inh{\nu F} = \nu X\ld \exists a\cln A\ld  \cons\;\pr{a, X^{B[a]}} && \axname{No Redundancy (Cojunk)}
\end{aligned}$
\item two destructors $\out, \nxt\in \Sigma$ together with the notations:\\
$~~~~\begin{aligned}
& x.\out \equiv \out\;x\\
& x.\nxt \equiv \nxt\;x
\end{aligned}$ 
\\
and the axioms:\\
$~~~~\begin{aligned}
&\forall a\cln A\ld \forall f\ld (\cons\;\pr{a, f}).\out = a && \axname{No Ambiguity (Coconfusion).1)}\\
&\forall a\cln A\ld \forall f\ld (\cons\;\pr{a, f}).\nxt = f && \axname{No Ambiguity (Coconfusion).2)}\\
&\forall x\cln \nu F\ld (\cons\;\pr{x.\out, x.\nxt}) = x && \axname{No Ambiguity (Coconfusion).3)}
\end{aligned}$
\end{itemize}

\noindent
\paragraph{Computing the greatest fixpoint} 
Since $\mathbf{gfp}(\phi_F)=\bigcap_{n\ge 0}\phi^n_F(M)=M\cap \phi_F(M)\cap \phi^2_F(M)\cap\cdots$, we have to investigate the infinite conjunction
\[
 \top \land \varphi_F(\top) \land \varphi^2_F(\top) \land \varphi^3_F(\top) \land \cdots \tag{\textsc{Gfp}}\label{eq:fin-ml}
\]
 in order to understand how the elements of the final coalgebra look like.
\begin{description}
\item[ $\varphi_F(\top)$.] We have $\cons\;\pr{a,\top^{B[a]}} = \exists x, y \ld x \land x.\out = a \land x.\nxt = y$.
\item[$\varphi_F^2(\top)$.] We have
\begin{align*}
\varphi_F^2(\top) & =\exists a\cln A\ld \cons\;\pr{a,\varphi_F(\top)^{B[a]}}\\
& = (\exists a\cln A\ld \exists x\ld \exists y\cln \sn{F\;\top} \ld x \land x.\out = a \land x.\nxt = y)\\
& = (\exists a,a'\cln A\ld \exists x, z\ld \exists y\cln \sn{F\;\top} \ld x \land x.\out = a \land x.\nxt = y\land y.\out = a' \land y.\nxt = z)
\end{align*}
\item [\ldots]
\end{description}

\begin{rem}
The ML pattern~\eqref{eq:fin-ml} can be seen as an ML informal translation of the colimit~\eqref{eq:fin}.
\end{rem}

\paragraph{Deriving Coinduction Principle}
Once we have seen how the greatest fixpoint is computed, we may derive the following \emph{Conduction Principle}:
\[
\dfrac{\forall a\cln A\ld \psi \limplies  \cons\;\pr{a,\psi} }{\inh{\psi \limplies \nu F}  }
\]
The justification for this principle is similar to that for streams given in~\cite{CLR21a}.

\section{Case Study: Lists Using Container Functors}
\label{sec:lists}

First, we express $L\; X = \objOne + E\times X$ as a unary container functor, using $\objOne \approxeq \sum_{\star\cln \objOne} X^{\objZero[\star]}$,  $E\approxeq\sum_{e\cln E}X^{\objZero[e]}$, and $X \approxeq \sum_{\star\cln \objOne}X^{\objOne[\star]}$:
\begin{align*}
L\; X & = \sum_{\star\cln \objOne} X^{\objZero[\star]} + \sum_{e\cln E}X^{\objZero[e]} \times \sum_{\star\cln \objOne}X^{\objOne[\star]}\\
&= \sum_{\star\cln \objOne} X^{\objZero[\star]} + \sum_{\pr{e,\star}\cln E\times \objOne}X^{\objZero[e] + \objOne[\star]}\\
&= \sum_{a\cln \objOne+E\times \objOne}X^{[\objZero, \langlebar \objZero, \objOne\ranglebar ][a]}
\end{align*}
Using the isomorphisms $E\times \objOne\approxeq E$ and $\langlebar \objZero, \objOne\ranglebar [\pr{e,\star}] = \objZero[e] + \objOne[\star] = \objZero + \objOne \approxeq \objOne$ in the category $\catC$, we obtain the reduced form $L^c\; X=\sum_{a\cln\objOne+E}X^{[\objZero,\objOne][a]}$ of the container functor $L$. From the definition of the sum of container functors we deduce that the elements of $\sum_{a\cln\objOne+E}X^{[\objZero,\objOne][a]}$ are pairs $\{\pr{\star, f} \mid f:\objZero\to X\}\uplus \{\pr{e, f'} \mid e:E, f':\objOne\to X\}$. The only function $\objZero\to X$ is $\initMorph$, and $f':\objOne\to X \approxeq f'\in X$.

The specification in ML of the initial algebra $\mu L^c$ includes:
\begin{itemize}
\item  a constructor symbol $\cons$ and a sort symbol $\mu F$;
\item the axioms:\\
$~~~~\begin{aligned}
& \exists y \cln \mu L^c\ld \cons\;\pr{\star, \initMorph} = y  && \axname{Functional.1}\\
& \forall e\cln E \ld \forall f'\cln \mu L^c \ld  \exists y\cln\mu L^c\ld \cons\;\pr{e, f'} = y   && \axname{Functional.2}\\
&\forall e \cln E\ld  \forall f'\cln \mu L^c \ld \cons\;\pr{\star, \initMorph} \not= \cons\;\pr{e, f'} && \axname{No Confusion.1}\\
&\forall e, e'\cln E\ld \forall f,f'\cln \mu L \ld  \cons\;\pr{e, f}=\cons\;\pr{e, f'}\limplies e = e' \land  f = f' && \axname{No Confusion.2}\\
&\inh{\mu L^c} = \mu X \ld (\cons\;\pr{\star, \initMorph} \lor  \cons\;\pr{E,X}) && \axname{No Junk}
\end{aligned}$
\end{itemize}
Comparing with the specification from ~\cite{CLR21a}, we obviously have the equivalences $\nil\approxeq \cons\;\pr{\star, \initMorph}$ and $\cons\;e\;\ell \approxeq \cons\;\pr{e, \ell}$.

The ML specification of the final coalgebra further includes:
\begin{itemize}
\item the destructor symbols $\out, \nxt\in \Sigma$;
\item the axioms\\
$~~~~\begin{aligned}
&\cons\;\pr{\star, \initMorph}\ld\out = \star && \axname{No Coconfusion.1.1)}\\
&\forall e\cln E \ld \forall f\cln \nu L^c \ld\cons\;\pr{e, f}\ld\out = e && \axname{No Coconfusion.1.2)}\\
&\cons\;\pr{\star, \initMorph}\ld\nxt = \initMorph && \axname{No Coconfusion.2.1)}\\
&\forall e\cln E \ld \forall f\cln \nu L^c \ld\cons\;\pr{e, f}\ld\nxt = f && \axname{No Coconfusion.2.2)}\\
&\forall x\cln \nu L^c\ld (\cons\;\pr{x.\out, x.\nxt}) = x && \axname{No Coconfusion.3)}\\
&\inh{\nu L^c} = \nu X \ld (\cons\;\pr{\star, \initMorph} \lor  \cons\;\pr{E,X}) && \axname{No Cojunk}
\end{aligned}$
\end{itemize}
Comparing with the specification of streams (inifinite lists) from ~\cite{CLR21a}, we obviously have the equivalences $\ell\ld\out \approxeq \hd\;\ell$ and $\ell\ld\nxt \approxeq \tl\;\ell$. Using  $\nil\approxeq \cons\;\pr{\star, \initMorph}$, we get $\hd\;\nil=\star$ and $\tl\;\nil=\initMorph$, which is different from the usual approach, where $\hd$ and $\tl$ are partial operations.

\section{Case Study: Moore Machines }
\label{sec:Moore}

Here we consider an example of (constant) exponential functor, whose ML specification is more tricky.
Moore machines (automata) have the signature given by the functor $M\;X=  O \times X^I$, where $O$ is for outputs and $I$ for inputs.

\subsubsection*{Capturing Final M-Coalgebra in ML Using Container Functors}

We first express $M$ as a unary container functor:
\begin{align*}
M\;X &= O\times X^I\\
&\approxeq \sum_{o\cln O}X^{\objZero} \times (\sum_{\star\cln \objOne}X^{\objOne})^{\sum_{i\cln I}X^{\objZero}}\\
&= \sum_{o\cln O}X^{\objZero} \times \sum_{g\cln  I\to \objOne}X^{\sum_{i\cln I}\objOne[g\;i]}\\
&=  \sum_{\pr{o,g}\cln (O\times (I\to \objOne))}X^{\objZero[o] + \sum_{i\cln I}\objOne[g\;i]}\\
&\approxeq \sum_{o\cln O}X^{\sum_{i\cln I}\objOne}\\
&\approxeq \sum_{o\cln O}X^{I}\\
&=M^c\;X
\end{align*}
The instantiation of the ML specification for the final coalgebra is as follows:
\begin{itemize}
\item the constructor $\cons\in\Sigma$ is specified by the following axioms:\\
$\begin{aligned}
&\forall o\cln O\ld \forall f\ld f\in X^{I}\limplies \exists x\ld x\in X\land\cons\;\pr{a,f} = x  && \axname{Functional}\\
&\forall o,o'\cln O\ld \forall f, f'\ld \cons\;\pr{a,f} = \cons\;\pr{a',f'} \limplies a = a' \land f = f'&& \axname{No Confusion}
\end{aligned}$
\item the destructors $\out, \nxt\in \Sigma$ are specified by the axioms:\\
$~~~~\begin{aligned}
&\forall o\cln O\ld \forall f\ld f\in X^{I} \limplies (\cons\;\pr{a, f}).\out = a && \axname{No Coconfusion.1)}\\
&\forall a\cln A\ld \forall f\ld f\ld f\in X^{I} \limplies (\cons\;\pr{a, f}).\nxt = f && \axname{No Coconfusion.2)}\\
&\forall x\cln \nu F\ld (\cons\;\pr{x.\out, x.\nxt}) = x && \axname{No Coconfusion.3)}
\end{aligned}$
\item the sort $\nu M^c$ with final semantics:\\
$~~~~\begin{aligned}
\inh{\nu M^c} = \nu X\ld \exists o\cln O\ld  \cons\;\pr{o, X^{I}} && \axname{No Cojunk}
\end{aligned}$

\end{itemize}
\noindent
We should have 
\[
\inh{\nu M^c} \approxeq \nu X\ld M^C\;X
\]
Computing $\top \land \varphi_M(\top) \land \varphi^2_M(\top) \land \varphi^3_M(\top) \land \cdots$, where $\varphi_M(X)\equiv  \exists o\cln O\ld \cons\;\pr{o,X^{I}}$:
\begin{description}
\item[$\varphi_M(\top)={}$]$\exists o_0\cln O\ld \cons\;\pr{o,X^{I}} = \exists o_0\cln O\ld \exists f_0\ld \cons\;\pr{o_0,f_0} \land (\forall i \cln I\ld \exists x_1\ld f_0\;i=x_1)$. When describing a dynamic system, the use of destructors is more intuitive: 
\[
\varphi_M(\top) = \exists x_0  \ld \exists o_0\cln O\ld x_0 \land (x_0.\out=o_0 \land \forall i\cln I\ld \exists y\ld x_0.\nxt\;i=y)
\]
It is easy to see that $x_0=\cons\;\pr{o_0, f_0}$ and $f_0\;i=x_0.\nxt\;i$.
\item[$\varphi^2_M(\top)={}$]$\exists o_1\cln O\ld \cons\;\pr{o_1,X^{I}}  = \exists o_1\cln O\ld \exists f_1\ld \cons\;\pr{o_1,f_1} \land (\forall i \cln I\ld \exists x_2\ld f_1\;i=x_2\land x_2\in \varphi_M(\top))$. Again, it becomes more suggestive using the destructors:
\begin{align*}
\varphi^2_M(\top) & =  \exists x_1  \ld \exists o_1\cln O\ld x_1 \land (x_1.\out=o_1 \land \forall i\cln I\ld \exists x_0 \ld x_1.\nxt\;i=x_0 \land x_0\in \varphi_M(\top))\\
& =  \exists x_1  \ld \exists o_1\cln O\ld x_1 \land (x_1.\out=o_1 \land \forall i\cln I\ld   \exists x_0  \ld \exists o_0\cln O\ld x_0.\out=o_0 \land \forall i\cln I\ld \exists y\ld x_0.\nxt\;i=y)
\end{align*}
\item[\ldots ] 
\end{description}
We have $\inh{\nu M}\ni x \approxeq f\in O^{I^*}$ iff $x.\out = f\;\epsilon$ and $\forall i\cln I\ld x.\nxt\;i=\overline{\mathit{tr}}(f)(i)$.
 
\section{Conclusion}
The technical experiments reported in this paper show that both the initial models and the final models for polynomial functors can be fully captured in matching logic (ML) using their representation as unary container functors. The ML specification of the polynomial functors is possible due to the fact the sum, product, and function sorts can be specified in ML~\cite{CLR21a}, and these specifications are a part of capturing the category of sets $\catC$ in ML.

A functor represented as a "classical" polynomial can be translated into a container functor shape using the fact the later are closed under sum, product, and exponential. However, the result could be cumbersome and  not easy to handle in matching logic because the construction starts from constant and identity functors. Therefore it is preferable to simplify it using the isomorphisms in the category of sets.

This result can help in defining in ML programming languages using both inductive data-types and coinductive data-types.
Another advantage is given by a better understanding of the abstract constructions from the category theory.
A possibly use is as follows:
\begin{itemize}
\item define in front-end a suitable syntax for data types intended to be defined as initial algebra or final coalgebra;
\item extract the canonical form of the functor underlying the front-end definition;
\item generate the corresponding ML theory;
\item derive the proof principles needed to soundly handle the defined data type.
\end{itemize}

Future work will focus on the following aspects of the proposed approach:
\begin{itemize}
\item a more formal presentation of the approach;
\item how to capture in ML the iteration principle and the primitive recursive principle;
\item extending the approach to larger classes of functors admitting initial algebras and final coalgebras, e.g., indexed containers~\cite{altenkirchGHMM15},  the bounded natural functors (BNFs) underlying Isabelle/HOL's datatypes ~\cite{TraytelPB12}, or the quotients of polynomial functors, experimentally implemented in Lean~\cite{AvigadCH19}.
\end{itemize}

\bibliographystyle{eptcs}
\bibliography{refs}

\begin{thebibliography}{10}
\providecommand{\bibitemdeclare}[2]{}
\providecommand{\surnamestart}{}
\providecommand{\surnameend}{}
\providecommand{\urlprefix}{Available at }
\providecommand{\url}[1]{\texttt{#1}}
\providecommand{\href}[2]{\texttt{#2}}
\providecommand{\urlalt}[2]{\href{#1}{#2}}
\providecommand{\doi}[1]{doi:\urlalt{https://doi.org/#1}{#1}}
\providecommand{\eprint}[1]{arXiv:\urlalt{https://arxiv.org/abs/#1}{#1}}
\providecommand{\bibinfo}[2]{#2}

\bibitemdeclare{article}{abbottAG05}
\bibitem{abbottAG05}
\bibinfo{author}{Michael~Gordon \surnamestart Abbott\surnameend},
  \bibinfo{author}{Thorsten \surnamestart Altenkirch\surnameend} \&
  \bibinfo{author}{Neil \surnamestart Ghani\surnameend} (\bibinfo{year}{2005}):
  \emph{\bibinfo{title}{Containers: Constructing strictly positive types}}.
\newblock {\slshape \bibinfo{journal}{Theor. Comput. Sci.}}
  \bibinfo{volume}{342}(\bibinfo{number}{1}), pp. \bibinfo{pages}{3--27},
  \doi{10.1016/j.tcs.2005.06.002}.

\bibitemdeclare{article}{Adamek03}
\bibitem{Adamek03}
\bibinfo{author}{Jir{\'{\i}} \surnamestart Ad{\'{a}}mek\surnameend}
  (\bibinfo{year}{2003}): \emph{\bibinfo{title}{On final coalgebras of
  continuous functors}}.
\newblock {\slshape \bibinfo{journal}{Theor. Comput. Sci.}}
  \bibinfo{volume}{294}(\bibinfo{number}{1/2}), pp. \bibinfo{pages}{3--29},
  \doi{10.1016/S0304-3975(01)00240-7}.

\bibitemdeclare{article}{altenkirchGHMM15}
\bibitem{altenkirchGHMM15}
\bibinfo{author}{Thorsten \surnamestart Altenkirch\surnameend},
  \bibinfo{author}{Neil \surnamestart Ghani\surnameend},
  \bibinfo{author}{Peter~G. \surnamestart Hancock\surnameend},
  \bibinfo{author}{Conor \surnamestart McBride\surnameend} \&
  \bibinfo{author}{Peter \surnamestart Morris\surnameend}
  (\bibinfo{year}{2015}): \emph{\bibinfo{title}{Indexed containers}}.
\newblock {\slshape \bibinfo{journal}{J. Funct. Program.}}
  \bibinfo{volume}{25}, \doi{10.1017/S095679681500009X}.

\bibitemdeclare{inproceedings}{altenkirchM09}
\bibitem{altenkirchM09}
\bibinfo{author}{Thorsten \surnamestart Altenkirch\surnameend} \&
  \bibinfo{author}{Peter \surnamestart Morris\surnameend}
  (\bibinfo{year}{2009}): \emph{\bibinfo{title}{Indexed Containers}}.
\newblock In: {\slshape \bibinfo{booktitle}{Proceedings of the 24th Annual
  {IEEE} Symposium on Logic in Computer Science, {LICS} 2009, 11-14 August
  2009, Los Angeles, CA, {USA}}}, \bibinfo{publisher}{{IEEE} Computer Society},
  pp. \bibinfo{pages}{277--285}, \doi{10.1109/LICS.2009.33}.

\bibitemdeclare{inproceedings}{AvigadCH19}
\bibitem{AvigadCH19}
\bibinfo{author}{Jeremy \surnamestart Avigad\surnameend},
  \bibinfo{author}{Mario \surnamestart Carneiro\surnameend} \&
  \bibinfo{author}{Simon \surnamestart Hudon\surnameend}
  (\bibinfo{year}{2019}): \emph{\bibinfo{title}{Data Types as Quotients of
  Polynomial Functors}}.
\newblock In \bibinfo{editor}{John \surnamestart Harrison\surnameend},
  \bibinfo{editor}{John \surnamestart O'Leary\surnameend} \&
  \bibinfo{editor}{Andrew \surnamestart Tolmach\surnameend}, editors: {\slshape
  \bibinfo{booktitle}{10th International Conference on Interactive Theorem
  Proving, {ITP} 2019, September 9-12, 2019, Portland, OR, {USA}}}, {\slshape
  \bibinfo{series}{LIPIcs}} \bibinfo{volume}{141}, \bibinfo{publisher}{Schloss
  Dagstuhl - Leibniz-Zentrum f{\"{u}}r Informatik}, pp.
  \bibinfo{pages}{6:1--6:19}, \doi{10.4230/LIPIcs.ITP.2019.6}.

\bibitemdeclare{techreport}{CLR20a}
\bibitem{CLR20a}
\bibinfo{author}{Xiaohong \surnamestart Chen\surnameend},
  \bibinfo{author}{Dorel \surnamestart Lucanu\surnameend} \&
  \bibinfo{author}{Grigore \surnamestart Ro\c{s}u\surnameend}
  (\bibinfo{year}{2020}): \emph{\bibinfo{title}{Initial Algebra Semantics in
  Matching Logic}}.
\newblock \bibinfo{type}{Technical Report}, \bibinfo{institution}{University of
  Illinois at Urbana-Champaign}.
\newblock \urlprefix\url{http://hdl.handle.net/2142/107781}.

\bibitemdeclare{article}{CLR21a}
\bibitem{CLR21a}
\bibinfo{author}{Xiaohong \surnamestart Chen\surnameend},
  \bibinfo{author}{Dorel \surnamestart Lucanu\surnameend} \&
  \bibinfo{author}{Grigore \surnamestart Ro\c{s}u\surnameend}
  (\bibinfo{year}{2021}): \emph{\bibinfo{title}{Matching logic explained}}.
\newblock {\slshape \bibinfo{journal}{Journal of Logical and Algebraic Methods
  in Programming}} \bibinfo{volume}{120}, pp. \bibinfo{pages}{1--36},
  \doi{10.1016/j.jlamp.2021.100638}.

\bibitemdeclare{inproceedings}{CR19}
\bibitem{CR19}
\bibinfo{author}{Xiaohong \surnamestart Chen\surnameend} \&
  \bibinfo{author}{Grigore \surnamestart Ro\c{s}u\surnameend}
  (\bibinfo{year}{2019}): \emph{\bibinfo{title}{Matching $\mu$-logic}}.
\newblock In: {\slshape \bibinfo{booktitle}{Proceedings of the
  34\textsuperscript{th} Annual ACM/IEEE Symposium on Logic in Computer Science
  (LICS'19)}}, \bibinfo{publisher}{IEEE}, \bibinfo{address}{Vancouver, Canada},
  pp. \bibinfo{pages}{1--13}, \doi{10.1109/LICS.2019.8785675}.

\bibitemdeclare{article}{ChenR20}
\bibitem{ChenR20}
\bibinfo{author}{Xiaohong \surnamestart Chen\surnameend} \&
  \bibinfo{author}{Grigore \surnamestart Roșu\surnameend}
  (\bibinfo{year}{2020}): \emph{\bibinfo{title}{A general approach to define
  binders using matching logic}}.
\newblock {\slshape \bibinfo{journal}{Proc. {ACM} Program. Lang.}}
  \bibinfo{volume}{4}(\bibinfo{number}{{ICFP}}), pp.
  \bibinfo{pages}{88:1--88:32}, \doi{10.1145/3408970}.

\bibitemdeclare{incollection}{jacobsRutten2012Introd}
\bibitem{jacobsRutten2012Introd}
\bibinfo{author}{Bart \surnamestart Jacobs\surnameend} \& \bibinfo{author}{Jan
  \surnamestart Rutten\surnameend} (\bibinfo{year}{2012}):
  \emph{\bibinfo{title}{An introduction to (co)algebra and (co)induction}}.
\newblock In \bibinfo{editor}{Davide \surnamestart Sangiorgi\surnameend} \&
  \bibinfo{editor}{Jan J. M.~M. \surnamestart Rutten\surnameend}, editors:
  {\slshape \bibinfo{booktitle}{Advanced Topics in Bisimulation and
  Coinduction}}, {\slshape \bibinfo{series}{Cambridge tracts in theoretical
  computer science}}~\bibinfo{volume}{52}, \bibinfo{publisher}{Cambridge
  University Press}, pp. \bibinfo{pages}{38--99}.
\newblock
  \urlprefix\url{http://www.cambridge.org/gb/knowledge/isbn/item6542021}.

\bibitemdeclare{article}{LinCTWR23}
\bibitem{LinCTWR23}
\bibinfo{author}{Zhengyao \surnamestart Lin\surnameend},
  \bibinfo{author}{Xiaohong \surnamestart Chen\surnameend},
  \bibinfo{author}{Minh{-}Thai \surnamestart Trinh\surnameend},
  \bibinfo{author}{John \surnamestart Wang\surnameend} \&
  \bibinfo{author}{Grigore \surnamestart Roșu\surnameend}
  (\bibinfo{year}{2023}): \emph{\bibinfo{title}{Generating Proof Certificates
  for a Language-Agnostic Deductive Program Verifier}}.
\newblock {\slshape \bibinfo{journal}{Proc. {ACM} Program. Lang.}}
  \bibinfo{volume}{7}(\bibinfo{number}{{OOPSLA1}}), pp.
  \bibinfo{pages}{56--84}, \doi{10.1145/3586029}.

\bibitemdeclare{article}{Ros17}
\bibitem{Ros17}
\bibinfo{author}{Grigore \surnamestart Ro\c{s}u\surnameend}
  (\bibinfo{year}{2017}): \emph{\bibinfo{title}{Matching logic}}.
\newblock {\slshape \bibinfo{journal}{Logical Methods in Computer Science}}
  \bibinfo{volume}{13}(\bibinfo{number}{4}), pp. \bibinfo{pages}{1--61},
  \doi{10.23638/LMCS-13(4:28)2017}.

\bibitemdeclare{inproceedings}{Rosu020}
\bibitem{Rosu020}
\bibinfo{author}{Grigore \surnamestart Roșu\surnameend} \&
  \bibinfo{author}{Xiaohong \surnamestart Chen\surnameend}
  (\bibinfo{year}{2020}): \emph{\bibinfo{title}{Matching logic: the foundation
  of the {K} framework (invited talk)}}.
\newblock In \bibinfo{editor}{Jasmin \surnamestart Blanchette\surnameend} \&
  \bibinfo{editor}{Catalin \surnamestart Hritcu\surnameend}, editors: {\slshape
  \bibinfo{booktitle}{Proceedings of the 9th {ACM} {SIGPLAN} International
  Conference on Certified Programs and Proofs, {CPP} 2020, New Orleans, LA,
  USA, January 20-21, 2020}}, \bibinfo{publisher}{{ACM}},
  p.~\bibinfo{pages}{1}, \doi{10.1145/3372885.3378574}.

\bibitemdeclare{article}{rutten00}
\bibitem{rutten00}
\bibinfo{author}{Jan J. M.~M. \surnamestart Rutten\surnameend}
  (\bibinfo{year}{2000}): \emph{\bibinfo{title}{Universal coalgebra: a theory
  of systems}}.
\newblock {\slshape \bibinfo{journal}{Theor. Comput. Sci.}}
  \bibinfo{volume}{249}(\bibinfo{number}{1}), pp. \bibinfo{pages}{3--80},
  \doi{10.1016/S0304-3975(00)00056-6}.

\bibitemdeclare{book}{sangiorgi-bisim-coind}
\bibitem{sangiorgi-bisim-coind}
\bibinfo{author}{Davide \surnamestart Sangiorgi\surnameend}
  (\bibinfo{year}{2012}): \emph{\bibinfo{title}{An Introduction to Bisimulation
  and Coinduction}}.
\newblock \bibinfo{publisher}{Cambridge University Press}.
\newblock \bibinfo{note}{A preliminary version of Chapter 2 from which we adapt
  paterial can be found at
  \url{http://www.cs.unibo.it/~sangio/DOC_public/corsoFL.pdf}}.

\bibitemdeclare{inproceedings}{TraytelPB12}
\bibitem{TraytelPB12}
\bibinfo{author}{Dmitriy \surnamestart Traytel\surnameend},
  \bibinfo{author}{Andrei \surnamestart Popescu\surnameend} \&
  \bibinfo{author}{Jasmin~Christian \surnamestart Blanchette\surnameend}
  (\bibinfo{year}{2012}): \emph{\bibinfo{title}{Foundational, Compositional
  (Co)datatypes for Higher-Order Logic: Category Theory Applied to Theorem
  Proving}}.
\newblock In: {\slshape \bibinfo{booktitle}{Proceedings of the 27th Annual
  {IEEE} Symposium on Logic in Computer Science, {LICS} 2012, Dubrovnik,
  Croatia, June 25-28, 2012}}, \bibinfo{publisher}{{IEEE} Computer Society},
  pp. \bibinfo{pages}{596--605}, \doi{10.1109/LICS.2012.75}.

\end{thebibliography}

\appendix

\section{Capturing Sort Operator in ML}
\label{ap:sorts}

We recall from~\cite{CLR21a} the ML specifications for the main operators over sorts.

\subsection{Sum Sort}
\label{sec:prod}
Given two sorts $s_1$ and $s_2$, we define a new sort $s_1{\otimes} s_2$,
called the \emph{product (sort) of $s_1$ and $s_2$},  as follows:
\begin{lstlisting}[language=matchinglogic]
spec $\textsf{SUM}_{s_1,s_2}$
  (*\Impts*) $\SORTS$
  (*\Smbs*) $\oplus, \iota_1, \iota_2,\epsilon_1,\epsilon_2$
  (*\Sugar*) $s_1{\oplus}s_2\equiv {\oplus}\,s_1\,s_2$
  (*\Axms*)
    $\prule{Sum Sort}$
      $s_1 \in \inh{\Sorts} \wedge s_2 \in \inh{\Sorts} \limplies
    s_1{\oplus}s_2\in\inh{\Sorts}$
    $\prule{Inject Left}$
      $\iota_1:{s_1} \to {s_1{\oplus} s_2}$
    $\prule{Inject Right}$
      $\iota_2:{s_2} \to {s_1{\oplus}s_2}$
    $\prule{Eject Left}$
      $\epsilon_1:{s_1{\oplus} s_2}\pto{s_1}$
    $\prule{Eject Right}$
      $\epsilon_2:{s_1{\oplus} s_2}\pto{s_2}$
    $\prule{Inverse InjEj1}$
      $\forall x\cln s_i\ld \epsilon_i\,(\iota_i\,x)=x,~i=1,2$
    $\prule{Inverse InjEj2}$
      $\forall x\cln s_{3-i}\ld \epsilon_i\,(\iota_{3-i}\,x)=\bot,~i=1,2$
    $\prule{CoProduct}$
      $\forall s_1,s_2\cln\Sorts\ld
      \inh{s_1{\oplus}s_2}\subseteq(\iota_1\,\inh{s_1})\lor(\iota_2\,\inh{s_2})$
endspec\end{lstlisting}
\subsection{Pair Sort}

\begin{lstlisting}[language=matchinglogic]
spec $\PAIR_{s_1,s_2}$ (*\Impts*) $\INH$                
(*\Smbs*) $\mathit{Pair},\mathit{pair},  \mathit{fst},   \mathit{snd}$
(*\Sugar*) $s_1 \otimes s_2 \equiv \mathit{Pair} \ s_1 \ s_2,
$ $\pr{\varphi_1,\varphi_2} \equiv \mathit{pair}\ \varphi_1 
\ \varphi_2$
(*\Axms*) (*$\sslash$ all axioms are quantified by ``$\forall s_1 , s_2 \cln 
\Sorts $''*)
$\begin{array}{ll}
\prule{Pair Sort} & (s_1 \otimes s_2) \cln \Sort \\
\prule{Pair} & 
\forall x_1 \cln s_1 \ld \forall x_2 \cln s_2 \ld 
\pr{x_1,x_2} \cln (s_1 \otimes s_2) \\
\prule{Pair Fst} &
\forall x_1 \cln s_1 \ld \forall x_2 \cln s_2 \ld
\mathit{fst} \ \pr{x_1,x_2} = x_1 \\
\prule{Pair Snd} &
\forall x_1 \cln s_1 \ld \forall x_2 \cln s_2 \ld
\mathit{snd} \ \pr{x_1,x_2} = x_2 \\
\prule{Pair Inj} & 
\forall x_1,y_1 \cln s_1 \ld \forall x_2,y_2 \cln s_2 \ld \\&\quad
\pr{x_1,x_2} {=} \pr{y_1,y_2} \limplies x_1 {=} x_2 \land y_1 {=} y_2 \\
\prule{Pair Domain} &  \inh{s_1 \otimes 
s_2} = \pr{\inh{s_1}, \inh{s_2}} 
\end{array}$
endspec 
\end{lstlisting}

\subsection{Function Sort}
\label{ap:func-sort}

\begin{lstlisting}[language=matchinglogic]
spec $\FUN_{s_1,s_2}$ (*\Impts*) $\INH$     
(*\Smbs*) $\mathit{Function}$
(*\Sugar*)   $s_1 \oto s_2 \equiv \mathit{Function} \ s_1 \ s_2$
(*\Axms*) (*// all axioms are quantified by ``$\forall s_1 , s_2 \cln 
\Sorts $''*)
$\begin{array}{ll}
\prule{Func Sort} & (s_1 \oto s_2) \cln \Sort \\
\prule{Func Domain} \!\!\!& 
\inh{s_1 \soto s_2} = \exists f \ld f \land \forall x \cln s_1 \ld 
(f \, x) \cln s_2 \\
\prule{Func Ext} & 
\forall f,g \cln s_1 {\oto} s_2 \ld
(\forall x \cln s_1 \ld f \ x = g \ x) \,{\limplies}\, f \,{=}\, g
\end{array}$
endspec
\end{lstlisting}

\begin{rem}
Note that we do not explicitly define the constructors of function sorts.
Instead, we axiomatize the \emph{behaviors} of a function.
Indeed, axiom \prule{Func Domain} states that a ``function'' $f$ of sort $s_1 
\oto s_2$ is one such that for any $x$ of sort $s_1$, $(f \ x)$ has sort $s_2$. 
Axiom \prule{Func Ext} states that two functions $f$ and $g$
of sort $s_1 \oto s_2$ are equal iff they are behavioral equivalent, i.e., 
they return the same values on all arguments of sort $s_1$.
\end{rem}

\end{document}